\documentclass{llncs}
\usepackage{psfig,amssymb}
\usepackage{eqnarray}

\begin{document}

\newcommand{\beq}{\begin{equation}}
\newcommand{\eeq}{\end{equation}}
\newcommand{\bea}{\begin{eqnarray*}}
\newcommand{\eea}{\end{eqnarray*}}
\newcommand{\eps}{\epsilon}
\newcommand{\ord}{{\cal O}}
\newcommand{\PP}{{\bf P}}
\newcommand{\CC}{{\bf CC}}
\newcommand{\NC}{{\bf NC}}
\newcommand{\NP}{{\bf NP}}
\newcommand{\NL}{{\bf NLOGSPACE}}
\newcommand{\LOGSPACE}{{\bf LOGSPACE}}
\newcommand{\Z}{{\Bbb Z}}
\newcommand{\avgr}{\overline{r}}
\newcommand{\act}{{\tt active}}
\newcommand{\occ}{{\tt occupied}}
\newcommand{\ot}{\overline{t}}
\newcommand{\arrow}[1]{\stackrel{#1}{\longrightarrow}}
\newcommand{\tarrow}[2]{\MYstack{\stackrel{#1}{\longrightarrow}}{#2}}

\def\MYstack#1#2{\begingroup 
  \ifinner\def\STYLE{\textstyle}\else\def\STYLE{\displaystyle}\fi
  \setbox0=\hbox{$\STYLE #1$} \dimen0=\ht0
  \setbox0=\hbox{$\stackrel{\STYLE #1}{\scriptstyle #2}$}
  \advance\dimen0 by -\ht0\relax
  \raisebox{\dimen0}{\box0}\endgroup
}

\title{Internal Diffusion-Limited Aggregation: \\
Parallel Algorithms and Complexity}
\author{Cristopher Moore \inst{1} and Jonathan Machta \inst{2}}
\institute{Santa Fe Institute \\ 1399 Hyde Park Road, Santa Fe NM 87501 USA
\\ {\tt moore@santafe.edu} \and
Department of Physics and Astronomy, University of Massachusetts\\
Amherst, MA 01003 USA
\\ {\tt machta@physics.umass.edu}}
\maketitle

\begin{abstract} The computational complexity of internal
diffusion-limited aggregation (DLA) is examined from both a
theoretical and a practical point of view.  We show that for two or
more dimensions, the problem of predicting the cluster from a given
set of paths is complete for the complexity class $\CC$, the subset of
$\PP$ characterized by circuits composed of comparator gates.
$\CC$-completeness is believed to imply that, in the worst case,
growing a cluster of size $n$ requires polynomial time in $n$ even on
a parallel computer.

\hspace*{0.25in} A parallel relaxation algorithm is presented that
uses the fact that clusters are nearly spherical to guess the cluster
from a given set of paths, and then corrects defects in the guessed
cluster through a non-local annihilation process.  The parallel
running time of the relaxation algorithm for two-dimensional internal
DLA is studied by simulating it on a serial computer.  The numerical
results are compatible with a running time that is either
polylogarithmic in $n$ or a small power of $n$.  Thus the
computational resources needed to grow large clusters are
significantly less on average than the worst-case analysis would
suggest.

\hspace*{0.25in} For a parallel machine with $k$ processors, we show
that random clusters in $d$ dimensions can be generated in $\ord((n/k
\,+\, \log k) \,n^{2/d})$ steps.  This is a significant speedup over
explicit sequential simulation, which takes $\ord(n^{1+2/d})$ time on
average.

\hspace*{0.25in} Finally, we show that in one dimension internal DLA
can be predicted in $\ord(\log n)$ parallel time, and so is in the
complexity class $\NC$.
\end{abstract}

\section{Introduction}

Internal diffusion-limited aggregation (DLA) is a cluster growth
process in which particles start at one or more sources within a
cluster, diffuse outward, and are added to the cluster at the first
site outside it they reach \cite{lawler1}.  By reversing figure and
ground, we can see this as a hole being hollowed out by particles
which remove sites from a surrounding material; therefore, this
process is sometimes called anti-DLA or diffusion-limited erosion
\cite{meakin,krug} and has been used to understand electrochemical
polishing.  Internal DLA is also equivalent to the problem of a
viscous fluid displacing an inviscid one in a porous medium
\cite{paterson,tang}.  If we add particles at a finite rate, rather
than one at a time, Gravner and Quastel~\cite{gravner} prove that the
hydrodynamic limit is the one-phase Stefan problem \cite{meirmanov},
which has been used as a model of a solid melting around a heat
source.  The purpose of this paper is to explore the computational
complexity of simulating internal DLA.

Internal DLA has quite different properties from its better known
cousin, ordinary DLA \cite{witten}, in which particles diffuse in from
infinity until they touch, and stick to, a growing cluster of sites.
Clusters grown in this way have a dendritic structure, and have been
used to model dielectric breakdown \cite{niemeyer}, electrochemical
deposition \cite{brady}, viscous fingering \cite{nittmann1}, snowflake
growth \cite{nittmann2}, the growth of vascular networks
\cite{family}, watershed formation \cite{masek}, neuron growth
\cite{hentschel}, and other phenomena.

\begin{figure}
\centerline{\psfig{file=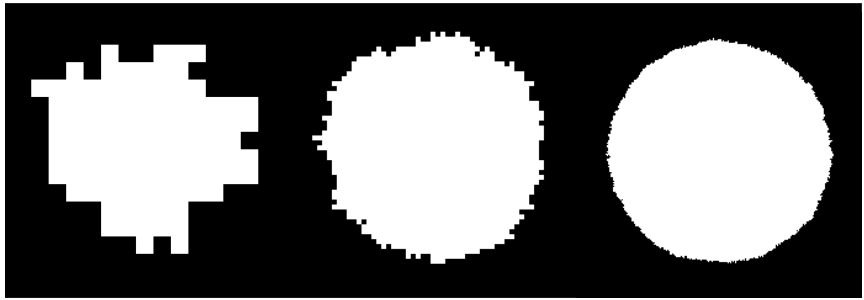,width=4.5in}}
\caption{Internal DLA clusters with 100, 1600, and 25600 particles.
Unlike ordinary DLA clusters, these have a circular shape.}
\label{clusterfig}
\end{figure}

While DLA tends to amplify irregularities in the cluster's boundary,
internal DLA tends to smooth them out.  For instance, in
Figure~\ref{clusterfig} we show a growing cluster at size 100, 1600,
and 25600, and it is clearly tending to a circular shape.  Lawler,
Bramson and Griffeath \cite{lawler1} showed in any number of
dimensions that the asymptotic shape of an internal DLA cluster with a
single source at the origin is spherical.  Formally, let $A_d r^d$ be
the volume of a $d$-dimensional ball of radius $r$.  Then they showed
that with probability 1, for any $\eps > 0$, the cluster with $A_d
r^d$ particles contains the ball of radius $r(1-\eps)$ centered on the
origin, and is contained within the ball of radius $r(1+\eps)$, for
sufficiently large $r$.

We can ask what the fluctuations in the boundary are, and define an
roughness or interface width $\xi$ where $\xi^2 = \langle (r-\avgr)^2
\rangle$.  Lawler \cite{lawler2} showed that in two or more
dimensions, $\xi$ scales at most as $r^{1/3}$ up to logarithmic
corrections.  For $d = 1$ the probability distribution of clusters can
be solved exactly \cite{lawler1}, and $\xi \sim r^{1/2}$ for clusters
of size $n = 2r$.

Krug and Meakin \cite{krug} have studied anti-DLA interfaces using
non-rigorous but presumably exact methods.  Their theory applies to a
line of sources and an asymptotically flat interface, but their
results should also apply to the point source and spherical interface
of internal DLA. They show that the interface width $\xi$ scales with
the length of the interface $L$ as $\log^{1/2}L$ for $d=2$ and goes
to a constant value for $d > 2$.  Their results are supported by
two-dimensional numerical simulations.

\begin{figure}
\centerline{\psfig{file=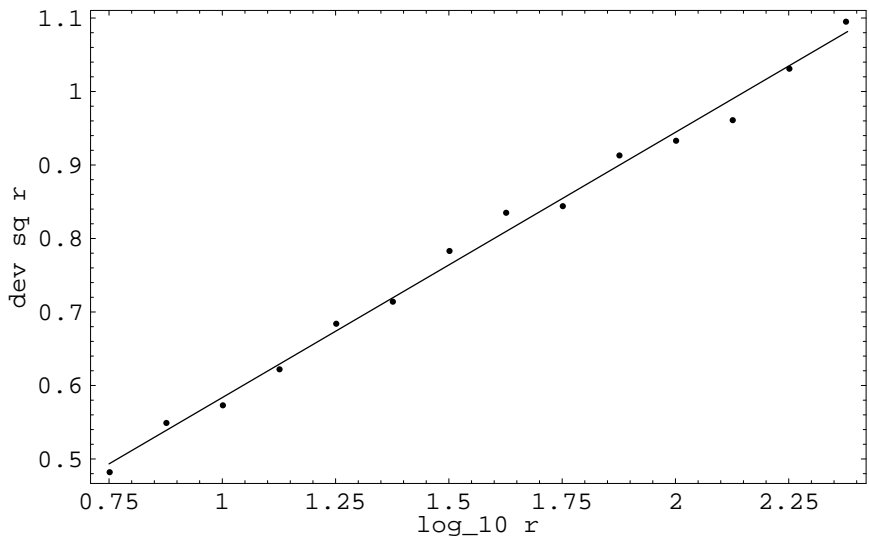,width=4.5in}}
\caption{A plot of the deviation $\langle (r-\avgr)^2 \rangle$ vs.\
$\log_{10} \avgr$, for $n$ up to $10^{5.25}$ averaged over $100$
trials each.  Deviations from circularity seem to grow only
logarithmically with $r$.}
\label{dev}
\end{figure}

We have performed simulations of internal DLA clusters in two
dimensions of size up to $n = 10^5$, with $100$ trials each.  The
average radius $\avgr$ of a point on the boundary converges very
quickly to $\sqrt{n/\pi}$, the radius of the circle with area $n$.  As
shown in Figure~\ref{dev}, the deviation $\xi^2$ seems to grow only
logarithmically with $r$, in agreement with Ref.~\cite{krug}.  Fitting
a plot of $\langle (r-\avgr)^2 \rangle$ vs.\ $\log_{10} \avgr$ gives a
slope of $0.36 \log_{10} \avgr = 0.16 \ln \avgr$.

\begin{figure}
\centerline{\psfig{file=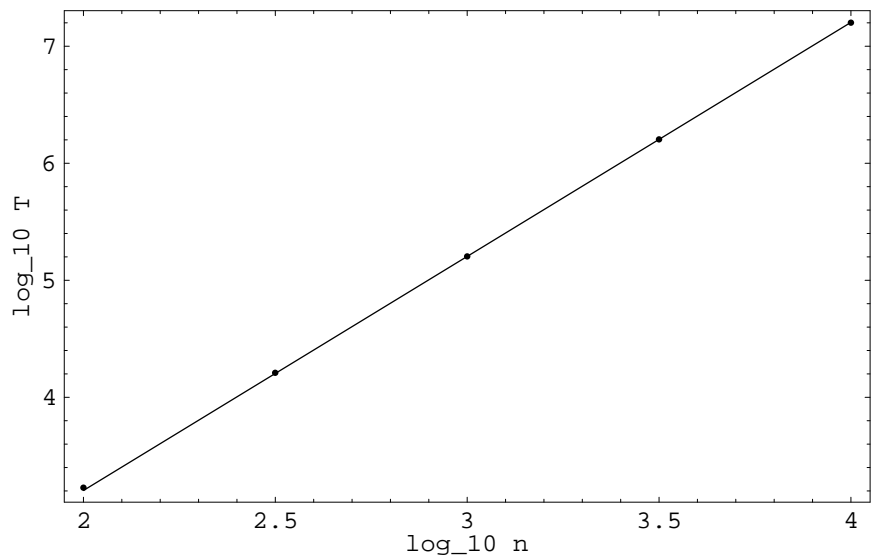,width=4.5in}}
\caption{A log-log plot of the computation time $T$ for explicit
simulation on a serial computer vs. the cluster size $n$ for $n$
ranging from $10^2$ to $10^4$, averaged over $100$ trials each.  The
straight line shows that the dynamical exponent is very close to $2$.}
\label{time}
\end{figure}

>From the scatter in Figure~\ref{dev}, it's clear that it would be nice
to have data for more trials and larger clusters.  However, since each
cluster has $n$ walks, and since each one has length proportional to
$r^2$ where $r \sim n^{1/d}$, the time it takes to explicitly simulate
the system on a serial computer is $T \sim n^z$ where the {\em
dynamical exponent} \cite{moriarty} $z = 1+2/d$.  In two dimensions,
$z = 2$ as shown in Figure~\ref{time}.  This places an upper limit on
the size of clusters we can generate, given limited computational
resources.


In this paper, we will discuss to what extent parallelization can help
us generate internal DLA clusters more quickly than explicit
simulation, both in the worst case and on average.  On the way, we
will show that the natural computational problem associated with
internal DLA is complete for a particular class of circuits, making it
one of the few known complete problems for this class.

We are interested in these questions for two reasons.  On a practical
level, to the extent that parallel computation becomes a reality, fast
algorithms will help us perform numerical experiments on larger
systems.  More philosophically, we believe that the computational
difficulty of predicting a system is a good measure of ``physical complexity'',
and that the complexity class a system belongs to says something
fundamental about its dynamics.  If a system is highly contingent on
its past, we have to simulate it explicitly, while if this dependence
is in some way sparse, we may be able to skip over much of its
history.  In our opinion, this is a fundamental distinction between
dynamical systems akin to integrability vs.\ chaos.  We hope that by
answering these questions for many systems, we will build a set of
intuitions about the relationships between complexity, dynamics, and
computation.

The paper is organized as follows.  In Section 2 we introduce the
basic notions of parallel complexity theory, including $\PP$, $\NC$,
and PRAMs.  In Section 3 we define comparator circuits and the class
$\CC$, and show that predicting internal DLA clusters is
$\CC$-complete.  Section 4 gives the most efficient algorithm we have
found for internal DLA, based on guessing the cluster shape and then
correcting this guess through a non-local annihilation process.

While this algorithm is very attractive, it requires a parallel
computer with a number of processors that grows polynomially in the
size of the cluster.  Using an equivalence between parallel and
sequential versions of internal DLA given in Section 5, in Section 6
we derive an efficient algorithm for the more realistic case in which
our computer has a fixed number of processors.  In Section 7 we show
that the one-dimensional case can be solved in logarithmic time, and
in Section 8 we conclude.  Finally, we give two additional algorithms
in the Appendix that may be of some interest.

\section{Prediction and computation}

Given a physical system, how much computational effort does it take to
predict it?  Must we simulate it step-by-step, or is it possible to
compress its history, and predict its behavior for $t$ time-steps on a
parallel computer with a computation time significantly less than $t$?
Computational complexity theory gives us a vocabulary to talk about
questions like these.

Computational complexity theory (see e.g.\ \cite{papa}) is the study
of the resources needed to solve problems and, more specifically, how
these resources increase as the problem size increases.  Computational
resources must be measured with respect to a specific model of
computation.  Happily, complexity theory is rather robust, in the
sense that superficially different models of computation lead to the
same hierarchy of complexity classes.  In this work we are primarily
interested in parallel computation, and the two models of parallel
computation that we use are families of Boolean circuits and parallel
random access machines (PRAMs).  A Boolean circuit is a feedforward
network of gates, typically AND, OR and NOT gates, although we will
also consider a more restricted set of gates below.  Boolean circuits
may be arranged in level such that all gates in a single level may be
evaluated simultaneously and the output of a given level is the input
of the next level.  Two primary complexity measures for a Boolean
circuits are {\em width} and {\em depth}.  Width is the largest number
of gates in a level and depth is the number of levels.  To solve a
problem of varying size, we need a family of circuits, with one
circuit for each problem size.

Two of the most important complexity classes are $\PP$ and $\NC$.
$\PP$ is the class of problems that can be solved by Boolean circuit
families where the circuit size is a polynomial of the problem size,
while $\NC$ is the subset of $\PP$ consisting of problems that can be
solved by families of circuits of polynomial size and polylogarithmic
depth.\footnote{A function $f(n)$ is polylogarithmic in $n$ if there
is a number $k$ such that $\lim_{n \rightarrow \infty} f(n)/\log^{k}n
=0$.}  $\PP$ is also the class of problems that can be solved in
polynomial time by a serial computer such as a Turing machine.  Within
$\NC$ are the nested subclasses $\NC^{k}$ of problems that can be
solved t polynomial size circuits of depth $\log^{k}n$ where $n$ is
the problem size.

The PRAM model of parallel computation is closer in design to real
parallel computers.  A PRAM is composed of many processor with
distince integer labels all connected to a shared memory.  Processors
run concurrently and all run the same program.  All processors can
read and write to a shared memory in unit time, an assumption that
cannot hold in the physical world as the number of processors is
scaled up.  Since each time step of a PRAM computation can be thought
of as a layer in a circuit which depends on the output of the previous
layer, the parallel time and the number of processors correspond
roughly to the depth and width of a circuit, respectively.  Thus $\NC$
is the set of problems that can be efficiently parallelized, i.e.\
solved in polylogarithmic time by a PRAM with a polynomial number of
processors.

Note that a PRAM requires a number of processors that grows with the
size of the problem, which may make it an impractical model of
parallel computation.  Below, we also discuss the more realistic case
where the number of processors is fixed.

Consider the following problem, called {\sc Circuit Value}: given a
description of a Boolean circuit composed of AND, OR and NOT gates,
and the truth values of the inputs, what is the truth value of the
output?  Clearly we can answer this by going through the circuit
layer-by-layer until we get to the output, so {\sc Circuit Value} is
in the class $\PP$.  In fact, it is the hardest such problem in the
sense that any other problem in $\PP$ can be reduced to it in a simple
way, and it is therefore {\em $\PP$-complete} \cite{greenlaw}.

The problem of computing the parity of $n$ bits, on the other hand,
can be solved in $\ord(\log n)$ parallel time.  Just XOR
pairs of bits, then pairs of pairs, and so on for $\lceil \log_2 n
\rceil$ steps.  This puts parity in the class $\NC^1$ of problems that
can be solved by a Boolean circuit of logarithmic depth and polynomial
(in this case, linear) width.

Just as computer scientists believe that $\NP$-complete problems
cannot be solved in polynomial time, they believe that $\PP$-complete
problems cannot be parallelized to polylogarithmic time.  If any
$\PP$-complete problem can be, then so can any problem in $\PP$, and
$\PP = \NC$, which would be almost as surprising as if $\NP = \PP$.
In other words, $\PP$-complete problems are believed to be {\em
inherently sequential}, so that much of the work has to be done
step-by-step, and even polynomially many processors cannot speed up
the computation very much \cite{greenlaw}.

In fact, predicting a number of physical problems has been shown to be
$\PP$-complete, for $d \ge 3$ in some cases and $d \ge 2$ in others.
These include ordinary DLA and fluid invasion
\cite{dla1,dla3}, the Ising model \cite{dla3,voting}, sandpiles
\cite{sand}, FHP and HPP lattice gases \cite{lgas}, cellular automata
with local voting rules \cite{voting}, and simple deterministic growth
models \cite{griff}.  Greenlaw et al.  \cite{greenlaw} have pointed
out that predicting cellular automata is $\PP$-complete in general,
since cellular automata exist (e.g.  \cite{mats}) which can simulate
universal Turing machines.  On the other hand, $\NC$ algorithms exist
for Eden growth \cite{dla2}, the Lorentz lattice gas \cite{lorentz},
and cellular automata with certain algebraic properties
\cite{moore1,moore2}.

Even if a speedup to polylogarithmic time isn't possible, we might
still hope for a polynomial speedup --- predicting physical time $t$
in $\ord(t^\alpha)$ parallel time for some $\alpha < 1$.  For
instance, in Ref.~\cite{moriarty} it was shown that though ordinary
DLA is $\PP$-complete, on average it can be parallelized to
$\ord(n^\alpha)$ time where $\alpha$ is related to the cluster's
fractal dimension.  To explore these finer distinctions, Condon
\cite{condon} introduced the idea of {\em strict $\PP$-completeness},
which can be used to put a lower bound on $\alpha$ unless all problems
in $\PP$ have a polynomial speedup.  Moore and Nordahl \cite{lgas}
discussed the strict $\PP$-completeness of predicting lattice gases,
and the same analysis could be applied to many of the problems listed
above.

\section{Comparator circuits}

\begin{figure}
\centerline{\psfig{file=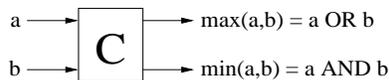,width=2in}}
\caption{Our notation for comparator gates.}
\label{comp}
\end{figure}

When discussing Boolean circuits, we usually take for granted that we
can {\em fan out} a wire by splitting it into as many copies as we
like.  This allows the output of one gate to be used as the input in
an arbitrary number of others.  Mayr and Subramanian \cite{mayr}
considered circuits where wires cannot be split except when this is
allowed explicitly by a gate, e.g.\ one with one input and two
outputs.  In particular, they considered the class $\CC$ of circuits
whose only gates are {\em comparators}, which have two inputs and two
outputs.  One output is the minimum (AND) of the inputs, and the other
is their maximum (OR).  We notate these as in Figure~\ref{comp}.

These circuits are incapable of fanout; in particular, they cannot
simulate, by restricting some of their initial values, a gate with
more non-constant outputs than inputs.  Mayr and Subramanian
\cite{mayr} show that this is true if and only if it is true for each
individual gate in the circuit, and they call this property being {\em
scatter-free}.  Because of this lack of fanout, the comparator circuit
value problem is believed not to be $\PP$-complete.  In fact, such
circuits can be evaluated fairly quickly, using an algorithm we will
now describe.

Comparator gates have the property that knowing either of their inputs
tells us what one of their outputs are.  If either input is 1, then
their maximum is 1, and if either input is 0, then their minimum is 0.
Moreover, in both cases the other output is simply the other input.
This means that any one of the $W$ inputs to a comparator circuit
determines the values of all the wires along some path connecting it
to one of its $W$ outputs, and the value of this output, leaving us
with a new comparator circuit of width $W-1$.  If the circuit has
depth $D$, this path can be found in $\ord(\log^2 D)$ parallel time,
since finding the transitive closure is in $\NC^2$ (or in fact in its
subset $\NL$ of nondeterministic logarithmic space \cite{papa}).

Repeating this algorithm for each input shows that comparator circuits
of width $W$ and depth $D$ can be evaluated in $\ord(W \,\log^2 D)$
parallel time.  More generally, by parallelizing the process of using
different inputs to simplify the circuit, Mayr and Subramanian showed
that a circuit of $N < WD$ gates can be evaluated in parallel time
$\ord(\min(W,D) \,\log^2 D) \,\lesssim\, \ord(\sqrt{N} \,\log^2 N)$,
since each simplification step reduces both the width and the depth by
at least one.  Thus a polynomial speedup to $N^{1/2}$ is always
possible.  On the other hand, there is no known algorithm for speeding
up the evaluation of comparator circuits to polylog time and it is
believed that $\NC$ and $\CC$ are incomparable.

We now show that $\CC$ circuits and internal DLA are intimately
linked.  Say that a particle is {\em active} if it is still moving
within the cluster, i.e.\ if it has not yet stuck to the outside of
the cluster because all the sites it has visited so far were already
occupied.  The input for our problem will be a list of moves $(t, i,
s)$, one for each time $0 \le t < T$, indicating that at time $t$
particle $i$, if it is still active, will visit site $s$.  Given such
a list, {\sc Internal DLA Prediction} is the problem of predicting the
set of occupied sites and the set of active particles at time $T$.
Note that this definition is quite general, allowing for arbitrary
topologies, multiple sources, and many particles moving at once.  Then
we have

\begin{proposition} {\sc Internal DLA Prediction} is in $\CC$.
\label{cc}
\end{proposition}

\begin{proof}  For each time $t$, define Boolean variables $\act_t(i)$
for each particle $i$ and $\occ_t(s)$ for each site $s$.  Then the
effect of a move $(t, i, s)$ is simply that of a comparator gate with
inputs $\act_t(i)$ and $\occ_t(s)$, and outputs $\act_{t+1}(i)$ and
$\occ_{t+1}(s)$:
\bea \occ_{t+1}(s) & = & \occ_t(s) \;\;\mbox{OR}\;\; \act_t(i) \\
     \act_{t+1}(i) & = & \occ_t(s) \;\;\mbox{AND}\;\; \act_t(i) \eea
This converts the list to a comparator circuit of size $T$ and width
$n+m$, where $n$ is the number of particles and $m$ is the total number
of sites named in the list.  The outputs $\occ_T(s)$ and $\act_T(i)$
give us the set of occupied sites and active particles at time $T$.
\qed \end{proof}

Conversely, any comparator circuit can be reduced to an internal DLA
problem on a square lattice with one particle at a time, of a size and
time polynomial in the size of the circuit.  Thus even this restricted
version of the problem is $\CC$-complete:

\begin{proposition}  {\sc Internal DLA Prediction} on a square lattice
is $\CC$-complete, even when restricted to one particle at a time.
\label{cccomp}
\end{proposition}

\begin{proof}  We will use sites of the cluster to store truth values,
with occupied and unoccupied sites representing true and false wires
respectively.  However, the same site will represent two different
wires at different times.  Our basic tool is the walk shown in
Figure~\ref{walk}, in which a particle comes from the origin and moves
down a horizontal conduit.  It steps off the conduit to visit site
$a$, continues to $b$ if $a$ is already occupied, and continues to a
previously unoccupied site $c$ if $b$ is already occupied.  If $t$ and
$t'$ are times before and after this walk, the effect on $\occ(a)$,
$\occ(b)$ and $\occ(c)$ is as follows:

\bea \occ_{t'}(a) & = & 1 \\
     \occ_{t'}(b) & = & \occ_t(a) \;\;\mbox{OR}\;\; \occ_t(b) \\
     \occ_{t'}(c) & = & \occ_t(a) \;\;\mbox{AND}\;\; \occ_t(b) \eea
Thus if the old values of $\occ(a)$ and $\occ(b)$ are the inputs to a
comparator gate, the new values of $\occ(b)$ and $\occ(c)$ are its outputs.

\begin{figure}
\centerline{\psfig{file=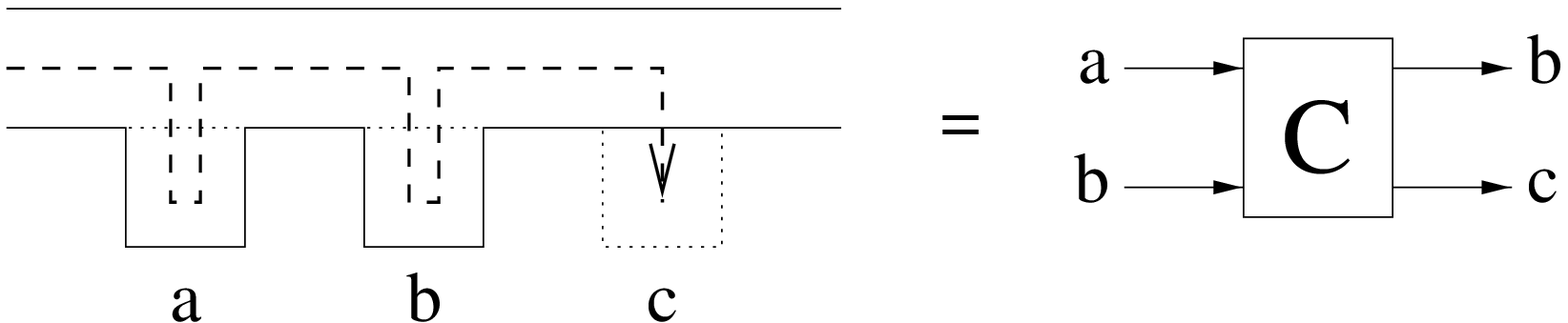,width=4in}}
\caption{A walk that implements a comparator gate.  After the walk,
$a$ will be occupied no matter what, $b$ will be occupied iff either
$a$ or $b$ were, and $c$ (which was unoccupied before) will be
occupied iff both $a$ and $b$ were.}
\label{walk}
\end{figure}

If a comparator circuit has $N$ gates, it has at most $4N$ wires,
which need at most $3N$ sites to represent their inputs and outputs.
If we place these sites contiguously along a row adjacent to the
conduit the particles use, and if the origin is at one end of this
conduit, each walk takes at most $3N + 6$ steps, and the total time
for $N$ such walks is $\ord(N^2)$.  \qed \end{proof}

Examining Propositions~\ref{cc} and \ref{cccomp}, we can see why
internal DLA is $\CC$-complete rather than $\PP$-complete.  While
sites can be used to store bits, these bits cannot be sensed by the
particles without being erased --- an unoccupied site becomes occupied
as soon as a particle touches it, and this particle then disappears.
Thus the system cannot make multiple copies of the truth value
represented by a site, and fanout is impossible.  In comparison to
this, collisions are more like NOR gates in ordinary DLA \cite{dla1},
and like Fredkin gates in the Reversible Aggregation model of D'Souza
and Margolus \cite{raissa}.

In addition to evaluating comparator circuits, other $\CC$-complete
problems include certain network stability problems and finding the
lexicographically first maximal matching in a graph \cite{mayr}.
While both $\NC$ and $\CC$ lie between $\NL$ and $\PP$, their apparent
incomparability suggests that parallelizability and a lack of fanout
are two very different properties.

Given Proposition~\ref{cc} and a supply of random bits, we can use
Mayr and Subramanian's algorithm to grow random clusters.
Specifically, for any $\eps > 0$ there is a parallel algorithm that
produces a random cluster of size $n$ in $d$ dimensions with
probability $1-\eps$ and runs in $\ord(n \,\log^d (n/\eps) \,\log^2
n)$ parallel time.  This is less efficient than the algorithm given in
the next section, but since the analysis is somewhat instructive, we
include it in Section~\ref{app1} of the Appendix.

\section{A parallel relaxation algorithm}
\label{secrelax}

In this section we describe a parallel relaxation algorithm for
generating internal DLA clusters.  The first step in the procedure is
to create an ordered list of the $n$ particles' random walks.  After
the walks are chosen, all we need to know is how far each particle
moves along its walk before it finds an unoccupied site and sticks
there.  We call this the {\em sticking point} or {\em label} of that
particle.

Call a configuration {\em well-ordered} if for every particle, there
are no labels of later particles along the path between the origin and
its label.  Call a configuration {\em singly-occupied} if no sites in
the cluster are empty or have more than one label, so that every
cluster site is the sticking point of exactly one particle.  There is
a unique well-ordered, singly-occupied configuration, and this
corresponds to the cluster that would have been produced by adding the
particles, one at a time according to the defining {\em sequential
dynamics}.  The idea of the relaxation algorithm is to start with a
reasonable initial configuration that is well-ordered but not
singly-occupied, and then to move particles' labels forward and
backward along their paths until it is singly-occupied as well.

Using the fact that clusters are very nearly spherical, it is easy to
create an initial configuration that is well-ordered and approximately
correct by placing the label of the $i^{\rm th}$ particle at the first
point on its walk where it reaches the radius of a sphere of volume
$i$.  This gives a spherical cluster of volume $n$ where some sites
are occupied by more than one label, and other sites have none.  We
refer to multiply occupied sites as {\em piles} of {\em pebbles}, one
pebble for each excess label, and unoccupied sites as {\em holes}.  We
then move the particle's labels in such a way that the number of
pebbles and holes decreases monotonically while keeping the
configuration well-ordered.

We begin with a description of the algorithm and its implementation on
a PRAM.  We then report on simulations of the algorithm that show
that its running time increases very slowly with the cluster size.

\subsection{Description of the relaxation algorithm}

The first step in the algorithm is to generate, in parallel, a list of
$n$ paths.  Path $i$ is an ordered list of distinct sites ${\bf
r}_i(1), {\bf r}_i(2), \ldots, {\bf r}_i(i)$, and corresponds to the
$i$'th walk in the sequential dynamics.  The sticking point or label
of path $i$ is at step $\tau_i$ and position ${\bf s}_i={\bf
r}_i(\tau_i)$.  Paths are constructed by generating random walks and
then compacting the walks to eliminate multiple visits to a site; this
can be done by a PRAM with a supply of random numbers in
polylogarithmic time.

To correctly simulate the sequential dynamics, the sticking points
must satisfy the property that for every path $i$ and for every time
$t \leq \tau_i$ there is exactly one path $j$ such that $j \leq i$ and
${\bf s}_j={\bf r}_i(t)$.  This insures that the walk $i$ arrives at
${\bf s}_i$ by moving within the already existing cluster and that no
two walks stick at the same site.  The well-ordering property is the
weaker property that, for all walks $i$ and $j$ and all $t < \tau_i$,
if ${\bf s}_j={\bf r}_i(t)$ then $j < i$.

Let us call the initial segment of a path up to and including its
label the {\em live} segment of the path.  A lattice site is live if
it is live for at least one path.  The cluster $S$ is the set of live
sites.  Note that this definition means the cluster may include
unoccupied sites, which we call holes.  The {\em perimeter} of the
cluster is the set of all sites that are not part of the cluster but
are neighbors of cluster sites.

The initial configuration of labels should be close to the typical
spherical configuration and must be well-ordered.  The expected radius
of the $i$'th walk's sticking point is $(i/A_d)^{1/d}$ where $A_d r^d$
is the volume of a sphere of sphere of radius $r$ in $d$ dimensions,
so we place the label for the $i$'th walk at the first site along it
whose distance from the origin is greater than this.  This can be
carried out in polylog time by a PRAM by calculating the radius of
each site, and ensures the well-ordering property as well.

To quantify the deviation from the correct configuration, we assign an
{\em energy} to a list of paths and their sticking points.  Let
$m({\bf r})$ be the number of labels at position ${\bf r}$ and let $S$
be the set of live positions (the cluster).  The energy $E$ is
\begin{equation}
E = \sum_{{\bf r} \in S} |m({\bf r})-1|
\end{equation}
Note that the correct configuration has energy zero and all other
well-ordered configurations have energy greater than zero.

The algorithm consists of moving labels forward and backward along the
walks.  Assuming that the current configuration is well-ordered, we
say that moving the label of walk $i$ from position ${\bf s}_i$ and
time $\tau_i$ to ${\bf s}^\prime_i$ and $\tau^\prime_i$ is {\em
allowed} if the resulting configuration is also well-ordered.

For every site in the cluster we will define a {\em hole index} and,
if the site is occupied, a {\em pebble index}.  The pebble index is
the highest label at a site, and the hole index is the lowest label of
all the walks that are live there.  At a site where $m({\bf r}) > 1$
is multiply-occupied, the $m({\bf r})-1$ excess labels there are
called {\em pebbles}, and the pebble index points to the pebble with
the highest label, i.e.\ the label of the last particle to stick at
the site in our current guess.  A site where $m({\bf r})=0$ is called
a {\em hole}, and the hole index tells us the first particle that
crosses it in our current guess.  Note that pebble and hole indices
are both defined at singly-occupied sites.

We will use two types of moves, pebble moves and hole moves.  A {\em
pebble move} consists of moving the pebble index at a given
multiply-occupied site outward along its
path until it reaches either (1) the perimeter of the cluster, (2) an
occupied site with an even higher pebble index, or (3) a hole with a
higher hole index.  We will call the first such site, moving outward
from its current position, its {\em destination}.  The destination is
the new sticking point for the particle.  The idea is that this
particle should not have stuck at its current site; since this site
was already occupied, it should have continued on to the first
unoccupied site on its path.

A pebble move preserves the well-ordering property, and does not
increase the energy.  If the destination is a hole, the pebble and the
hole are annihilated, and the energy decreases by two.  If the
destination is on the perimeter, the pebble is annihilated, a new site
is added at the perimeter, and the energy decreases by one.  Finally,
if the destination is a site with a higher pebble index the energy is
unchanged.

We can perform many pebble moves in parallel, by determining all the
pebbles' destinations and moving them there simultaneously.  This
might result in two particles being placed on the same site, but the
well-ordering property is still preserved, and the energy is never
increased.  In fact, we can do more than this in parallel.  If a
pebble's destination is a singly-occupied site with a higher pebble
index, a new pebble with that higher index is created by the move; the
new pebble, in turn, might have a singly-occupied destination with a
yet higher pebble index, and so on.  Thus a series of pebbles can
cascade outward until the last one falls in a hole or sticks at the
perimeter.

We can carry out an entire cascade of this kind in one polylog time,
parallel step.  For each occupied site there is a pebble index and a
destination site where that pebble index would move if the site were
multiply-occupied.  The directed bonds connecting the pebble indices
of occupied sites and their destinations form a directed forest of
potential pebble moves.  In a single {\em pebble sweep}, we move some
of the pebble indices in this forest to their destinations.  The
pebble indices that are moved are the ones that can be reached in this
directed forest from a multiply-occupied site and are thus part of the
cascade of pebble moves.  Determining the forest of pebble moves and
moving the pebbles to their destinations can be done in a $\ord(\log^2
n)$ time by a PRAM using graph reachability \cite{papa}.

A hole exists because one or more particles cross a site as if it were
occupied, even though no particle is said to stick there in our
current guess.  The hole index tells us the label of the first such
particle to do so.  Therefore, a hole move consists of moving that
particle's label inward along its path to fill the hole; since that site
was unoccupied when it got there, it should have stuck there instead.

A hole move is always allowed and does not increase the energy.  If
the moved label was at a multiply-occupied site the energy decreases
by two.  If the moved label was at a singly-occupied site, it creates
a new hole there, leaving the energy unchanged (since some other
particle relied on that site being occupied in order to cross it)
unless that site is just inside the perimeter, in which case it is
removed from the cluster and the energy decreases by one.

We will perform a {\em hole sweep} of many hole moves in parallel.  We
can have cascades of hole moves just as with pebble moves, in which a
hole created by moving a label from a singly-occupied site is filled
in turn by a particle from another singly-occupied site, and so on.
In general, there is a forest of hole moves where some particles are
indexed by the hole indices of more than one hole.  In this situation,
we move the particle to fill the hole at the earliest time along its
path, and the other holes go unfilled until a later sweep.

In each sweep of each kind, at least one pebble (the outermost along
its path) or at least one hole (the innermost) will be removed.  Since
there are no more than $n$ pebbles and holes in the initial
configuration, and since each sweep can be performed in
polylogarithmic time by a PRAM, the running time is no worse than
$\ord(n \log^k n)$.  However, since many pebbles and holes are
typically annihilated in a single sweep we expect much better
performance than this from the algorithm on average, and this is borne
out by the numerical results in the next section.

We conclude this section with a discussion of the processor
requirements of the algorithm.  The usual algorithm for graph
reachability, which we use to determine cascades of pebble and
hole moves, involves repeatedly squaring the adjacency matrix
of a directed graph, and takes $N^3$ processors on a graph of
size $N$.  Since the graph in question consists of the $n$ sites
of the cluster itself, our algorithm needs $\ord(n^3)$ processors to
carry out pebble or hole sweeps in polylogarithmic time.

\subsection{Simulations of the relaxation algorithm}

Our algorithm consists of alternating pebble sweeps and hole sweeps
until the energy is zero, at which point the cluster is in the correct
configuration.  How many steps are required to do this?  To explore
this question we simulated the relaxation algorithm on a serial
machine, and measured the average number of sweeps as a function of
system size.

The simulation is carried out using two data structures, one
representing the lattice sites and the other representing the walks.
Stored with each lattice site is its pebble index, hole index and
the total number of walks sticking at the site.

It would require $\ord(n^{2/d+1})$ memory, far too much, to store the
full trajectories of $n$ walks.  Therefore, we trade time for memory,
and define each walk by a four byte integer that is the seed for a
linear congruential random number generator.  The walk is generated as
needed using the random number generator initialized by this integer
{\em pathname}.  The linear congruential random number generator takes
an integer and produces a new integer.  Thus, the pathname of the walk
is a function of the step along the walk and the walk can be generated
outward from any point where the current pathname is known by
application of the random number generator.  The data stored for each
walk is its pathname at the origin and its pathname, time and position
at its current sticking point.  In addition, the pathname and time
associated with the hole label at each lattice site is stored with
each lattice site.

Given this data structure it is straightforward to simulate pebble and
hole sweeps without actually determining the forests of pebble and
hole moves.  For a pebble sweep, all the sites of the lattice are
visited in order.  If a site is multiply-occupied and the pebble index
of the site has not yet been moved, this label is moved outward along
its path to its destination, which may create a new multiply-occupied
site.  The algorithm cycles through the lattice until no further
pebbles are moved.  For a hole sweep, all the sites of the lattice are
visited in order.  If a site is a hole then the particle corresponding
to its hole index is moved to the site, which may create a new hole.
This process is continued until no further holes are moved.  After
each sweep we also update the pebble and hole indices of each site.
One step of the algorithm consists of a pebble sweep, an update of the
site information, a hole sweep and another update of the site
information.  A single step of this sequential simulation corresponds
to polylogarithmic parallel time on a PRAM --- however, given the
amount of effort to do all this, this is certainly not the best way to
grow internal DLA clusters on a serial computer!

\begin{figure}
\centerline{\psfig{file=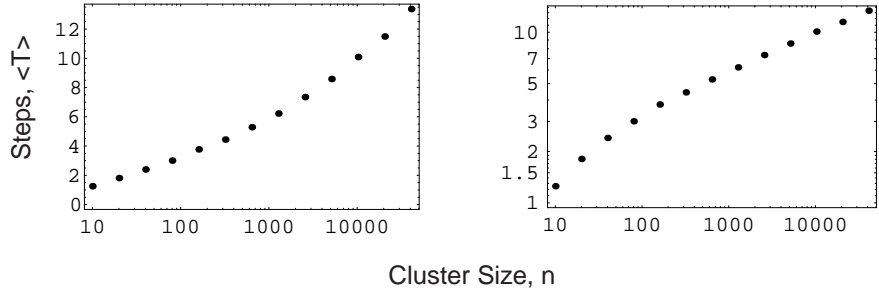,width=5in}}
\caption{The number of steps $\langle T \rangle$ for the relaxation
algorithm to converge vs.\ cluster size $n$ for $n$ in the range 10 to
40960. The left panel is a semi-log plot and the right panel is a
log-log plot. }
\label{jontime}
\end{figure}

We have run sequential simulations of the relaxation algorithm for a
series cluster sizes from 10 through 40960.  To check the algorithm,
we confirmed that the clusters obtained from the relaxation algorithm
are exactly those obtained from the sequential dynamics for the same
walks ordered by index.  We measured how the energy decays to zero as
a function of the number of steps and calculated the average number of
steps for the algorithm to converge as a function of cluster size.
Table~\ref{tab4} shows the average number of steps required by the
algorithm to reach the correct configuration as a function of the
cluster size $n$ and Figure~\ref{jontime} plots the data.  In the left
panel of the Figure, the data is presented as a semi-log plot and on
the right panel as a log-log plot.  Neither curve is straight, which
suggests a slowly varying function between $\log n$ and a power of
$n$.  The best fit for $n \geq 160$ to the form $a + b n^z$ yields
$z=0.18$.  For the same range of $n$, the best fit to the form $a + b
\log^\alpha n$ yields $\alpha = 1.6$.  Both fits are reasonably good
on this limited range of $n$, so we cannot say for sure whether the
asymptotic behavior is polynomial or polylogarithmic.  However, if it
is a polynomial, a power of $0.18$ is unusually small.  It should be
noted that even if the asymptotic behavior is polylogarithmic, the
actually running time on a PRAM would have an additional
polylogarithmic factor, giving a larger value of $\alpha$, since each
step of the algorithm itself requires polylogarithmic parallel time.

\begin{table}
\[ \begin{array}{||ll|ll||} \hline
n & \langle T \rangle& n & \langle T \rangle \\ \hline
10    & 1.25 & 1280  & 6.22 \\
20    & 1.81 & 2560  & 7.35 \\
40    & 2.40 & 5120  & 8.58 \\
80    & 3.01 & 10240 & 10.09 \\
160   & 3.77 & 20480 & 11.49 \\
320   & 4.45 & 40960 & 13.37 \\
640   & 5.29 & & \\ \hline

\end{array} \]
\caption{The average number of steps $\langle T \rangle$ for the
relaxation algorithm to reach the correct configuration versus cluster
size $n$.}
\label{tab4}
\end{table}

\begin{figure}
\centerline{\psfig{file=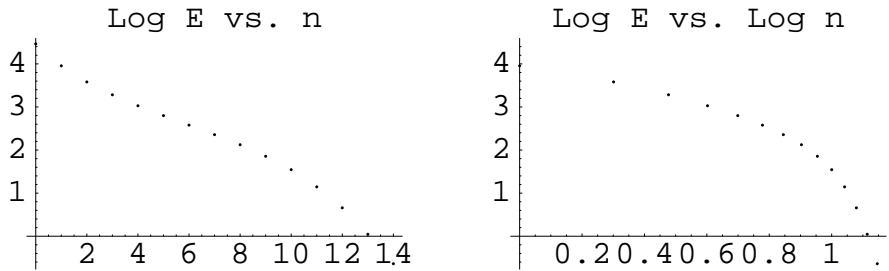,width=5in}}
\caption{The energy as a function of the number of steps, averaged
over 100 trials, for clusters of size $4 \cdot 10^4$, plotted both
semi-log (on the left) and log-log (on the right).  While the data
shows an s-curve, which in the text we argue shows three regimes of
the relaxation process, the relatively straight line on the left seems
to indicate that the energy is decreasing exponentially in the number
of steps.}
\label{enfig}
\end{figure}

\begin{figure}
\centerline{\psfig{file=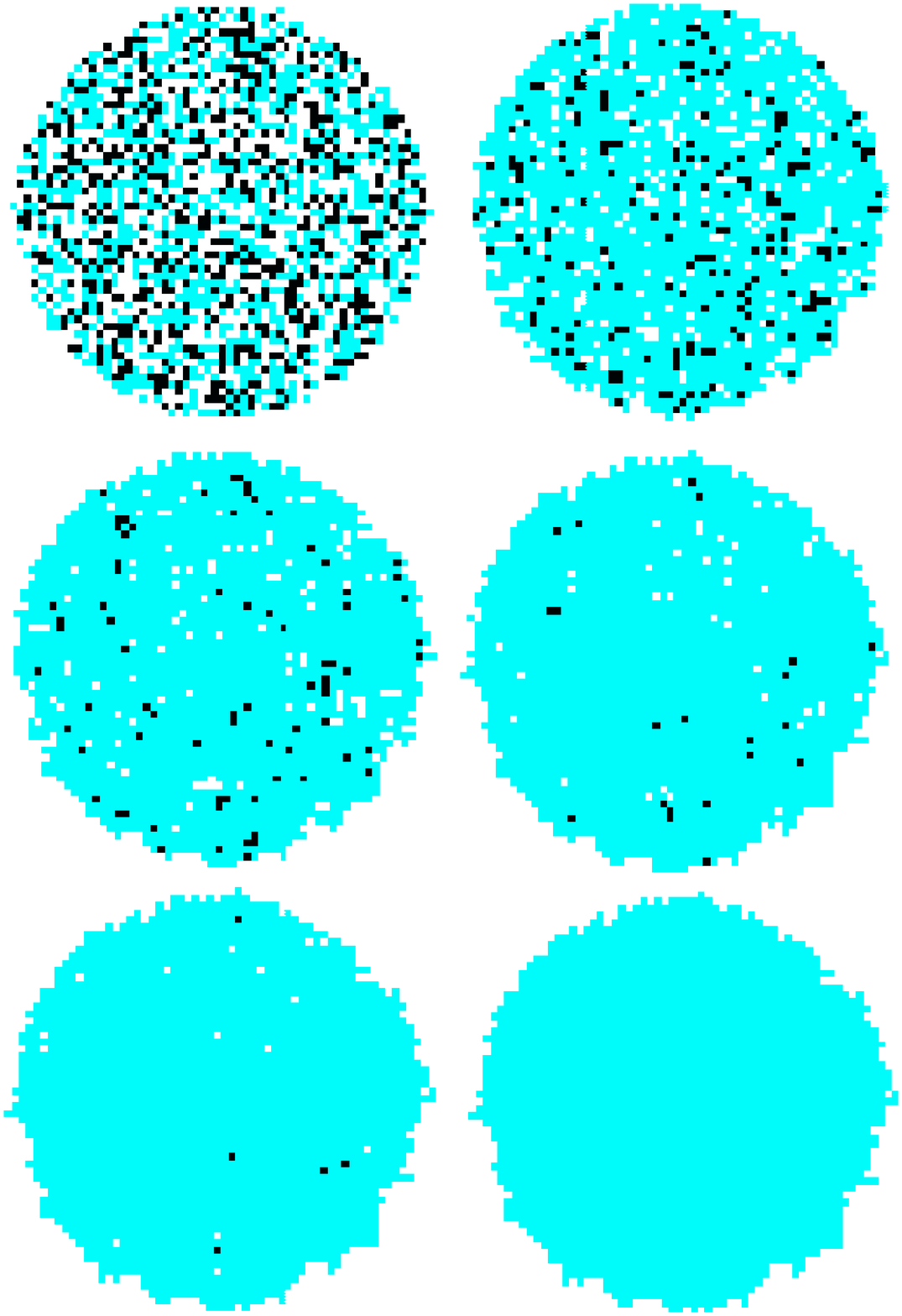,height=8in}}
\caption{A cluster of size 2500 after 0, 1, 2, 3, 4 steps and its
final shape after 6 steps.
Pebbles, holes and singly-occupied sites are black, white and grey,
respectively.}
\label{jonrev}
\end{figure}

Figure~\ref{enfig} shows the energy as a function of the number of
steps, averaged over 100 trials, for clusters of size $4 \cdot 10^4$.
The curve is close to a straight line on the semi-log plot but shows a
slight ``s'' shape, which we believe is related to the existence of
three regimes in the dynamics.  Figure~\ref{jonrev} shows snapshots of
the system as the algorithm converges to a correct cluster of size
2500.  In the first regime (steps 0-2 in Fig.~\ref{jonrev}) there is a
high density of pebbles and holes and these annihilate one another
locally, giving an exponential decrease in the energy.  In the middle
regime (step 3) pebbles and holes are separated into domains, and
annihilation occurs mainly at the boundaries between these and at the
perimeter of the cluster.  In the final regime (step 4) the energy
again decreases rapidly when almost all pebbles and holes are
independently annihilated at the perimeter in a few steps.

Because the middle regime is the slowest and does not become
well-developed until $n$ is large, it is difficult to extract the
asymptotic behavior of the running time of the algorithm from the
numerics even at relatively large values of $n$.  If this regime is
similar to local diffusion and annihilation processes in the plane
where particles of opposite type separate into domains, then we would
expect the energy to decrease as a power-law function of time
\cite{toussaint,bramson}.  However, since our moves are non-local the
particles don't have to take the time to diffuse to each other, and
the data in Figure~\ref{jontime} and Figure~\ref{enfig} suggests that
the decay may in fact be closer to exponential.

\section{Commutativity and parallel vs.\ sequential growth}
\label{seccomm}

While the algorithm of the previous section works very well on
massively parallel computers, we also want fast algorithms for the
more practical case where our parallel computer has a fixed number of
processors.  To do this, in this section we will show the surprising
fact that a wide variety of versions of internal DLA, ranging from
adding one particle at a time to adding them all at once, all produce
the same probability distribution of cluster shapes.

Diaconis and Fulton \cite{diaconis} showed that internal DLA has a
remarkable kind of commutativity.  If we have a probability
distribution $P$ of cluster shapes, and we define $T_x(P)$ as the new
distribution resulting from adding a particle with initial position
$x$, then $T_x(T_y(P)) = T_y(T_x(P))$ for any sites $x$ and $y$.  In
other words, if we add two particles with two initial positions, it
doesn't matter which order we add them in.

Their proof is quite general, and does not rely on the topology of
the lattice or any particular set of transition probabilities between
sites.  It relies on the fact that the particles only interact when
the one starting at $x$ is added to the cluster at a site $s$, and the
one starting at $y$ passes through $s$ to another site $t$.  The
probability of this is
\[ P(x \to s) \,P(y \to s) \,P(s \to t) \]
This is symmetric in $x$ and $y$, since the walk from $s$ to $t$ can
just as easily be taken by either particle once the other one has
occupied $s$.  This commutativity does not hold for standard DLA, on
the other hand, because particles block each others' paths rather than
facilitating them.

Closely related to commutativity is parallelizability.\footnote{Note
that we are using the word `parallel' in two different ways in this
paper: first, for the operation of algorithms on parallel computers,
and second, for the growth model where multiple particles are released
at once.}  If we start two particles at the same time and run them in
parallel, by the time the first one is deposited at a site $s$, the
other one will be exactly as likely to be at any given position as it
would be if it were released sequentially after the first one completed
its walk, with the one caveat that the number of steps it takes the
second particle to reach $s$ must be greater than or equal to that the
first particle took.  For pairs of walks where this inequality is
violated, we can swap these parts of the particles' walks.

To prove this formally, let $P(S+x+y)$ be the probability that adding
two particles at the origin increases a cluster $S$ by two sites $x$
and $y$.  Call $P(x \arrow{S} y)$ the probability that a particle
starting at $x$ sticks to a new site $y$, and $P(x \tarrow{S}{t} y)$
the probability that a particle starting at $x$ visits $y$ for the
first time after $t$ steps.  For bookkeeping purposes we will use
subscripts $P_1$ and $P_2$ to show which particle takes which path,
but of course the probability doesn't depend on this.

If we release the two particles sequentially, we have
\[ P(S+x+y) = P_1(0 \arrow{S} x) \,P_2(0 \arrow{S+x} y)
        \,+\, P_1(0 \arrow{S} y) \,P_2(0 \arrow{S+y} x) \]
Taking one of these and separating it into terms counting non-interacting
walks and interacting ones gives
\begin{equationarray*}{rclr}
P(S+x+y) & = & P_1(0 \arrow{S} x) \,P_2(0 \arrow{S} y)
           		                  & \mbox{(non-interacting)} \\
& + & \;P_1(0 \arrow{S} x) \,P_2(0 \arrow{S} x) \,P_2(x \arrow{S+x} y)
                                          & \mbox{(interacting)} \\
& + & \;(x \Leftrightarrow y) & \\
& = & 2 \,P(0 \arrow{S} x) \,P(0 \arrow{S} y) & \\
& + & \;P(0 \arrow{S} x)^2 \,P(x \arrow{S+x} y) & \\
& + & \;P(0 \arrow{S} y)^2 \,P(y \arrow{S+y} x)
\end{equationarray*}
(here $(x \Leftrightarrow y)$ indicates the corresponding terms with
$x$ and $y$ switched).

If instead we release the two particles at the same time, let's assume
that particle 1 sticks at $x$ and particle 2 sticks at $y$.  As
before, we separate $P(S+x+y)$ into an interacting and a
non-interacting part.  The interacting part can be divided into terms
depending on which particle sticks first.  In the first set of terms
particle 1 sticks at time $t_1$, and particle 2 first visits $x$ at
some time $t_2 \ge t_1$ and then travels from $x$ to $y$.  In the
second set of terms, particle 2 sticks at time $t_2$, and particle 1
first visits $y$ at some time $t_1 > t_2$ and then travels from $y$ to
$x$.  Note that $t_1=t_2$ is included in the first set of terms, since
if both particles reach $x$ at the same time, by convention we deposit
particle 1 and keep particle 2 active.

Then for the parallel case we have
\begin{equationarray*}{rclr}
P(S+x+y) & = & P_1(0 \arrow{S} x) \,P_2(0 \arrow{S} y)
			                & \mbox{(non-interacting)} \\
& + & \;\left( \sum_{t_1 \le t_2}
         P_1(0 \tarrow{S}{t_1} x) \,P_2(0 \tarrow{S}{t_2} x) \right)
      \,P_2(x \arrow{S+x} y)            & \mbox{(1 sticks first)} \\
& + & \;\left( \sum_{t_1 > t_2}
         P_1(0 \tarrow{S}{t_1} y) \,P_2(0 \tarrow{S}{t_2} y) \right)
      \,P_1(y \arrow{S+y} x)            & \mbox{(2 sticks first)} \\
& + & \;(x \Leftrightarrow y) & \\
& = & 2 \,P(0 \arrow{S} x) \,P(0 \arrow{S} y) & \\
& + & \;\left( \sum_{t_1 \le t_2} + \sum_{t_1 > t_2} \right)
      \left( P_1(0 \tarrow{S}{t_1} x) \,P_2(0 \tarrow{S}{t_2} x) \right)
      \,P_2(x \arrow{S+x} y) & \\
& + & \;\left( \sum_{t_1 \le t_2} + \sum_{t_1 > t_2} \right)
      \left( P_1(0 \tarrow{S}{t_1} y) \,P_2(0 \tarrow{S}{t_2} y) \right)
      \,P_1(y \arrow{S+y} x) & \\
& = & 2 \,P(0 \arrow{S} x) \,P(0 \arrow{S} y) & \\
& + & \;P(0 \arrow{S} x)^2 \,P(x \arrow{S+x} y) & \\
& + & \;P(0 \arrow{S} y)^2 \,P(y \arrow{S+y} x)
\end{equationarray*}
This is the same as the expression derived above for the sequential
case, and so we get the same probability distribution whether we
release the particles sequentially or in parallel.

Lawler, Bramson and Griffeath \cite{lawler1} give the following
general argument, which works for any number of particles.  Suppose we
choose a random walk for each particle in advance.  Each potential
site in the cluster is visited by many different particles.  We can
consider a variety of protocols for determining which particle visits
that site first and sticks there, while the other particles remain
active.  Some obvious protocols are
\begin{itemize}
\item Sequential growth, where each particle has an index indicating
the order in which it was released, and we attach the particle with
the lowest index.
\item Parallel growth, where we attach the particle that visits this
site earliest in its walk, using the index to break ties.
\item A mix of these, where particles are released in a series of
waves or at various times.
\end{itemize}
Each such protocol defines a growth model, and {\em all} such models
are equivalent, as long as these protocols depend only on the past,
i.e.\ on the part of the particles' walks that precedes their visit to
the site in question.  There are two main ingredients to the proof.
First, if the protocol depends only on the past then the future of
each particle's walk is free of correlations with the fate of the
others.  Second, past sections of different particles' walks can be
swapped with each other as we did in the two-particle case to
transform a run under one protocol into a run under another.

One such protocol, which adds a shell of constant thickness to the
cluster at each step, leads to a reasonably fast parallel algorithm.
It requires $\ord(n^{1+2/d})$ processors, and grows random clusters of
size $n$ in $d$ dimensions in time $\ord(n^{1/d} \,\log n)$.  While it
is inferior to the algorithm of Section~\ref{secrelax}, it is
conceptually much simpler.  We give it in Section~\ref{app2} of the
Appendix.

\section{Practical parallelism: a fixed number of processors}

While the parallel algorithms given in Section~\ref{secrelax} and the
Appendix are interesting, they are impractical given the current state
of parallel computing technology.  They rely on a polynomially growing
number of processors, all of which have access to a shared memory.
Communication delays make it difficult to build shared-memory machines
with many processors, as opposed to distributed-memory machines where
each processor has a local cache.  To date the largest shared-memory
computers have 16 processors, although computers that simulate shared
memory with a nonuniform cost for access have been built with many
more.  At the time of this writing, the largest CC-NUMA (cache
coherent non-uniform memory access) computer is ASCI Blue Mountain at
Los Alamos, with 6144 processors.

In this section, we ask a more practical question: how much can we
speed up an internal DLA simulation, specifically for generating
random clusters, if we have a shared-memory computer with $k$
processors?  We will assume we have a {\em concurrent-read, priority
concurrent-write} (CRCW) machine.  In a CRCW PRAM, each processor has
an index.  Two or more processors can read the same bit from memory
simultaneously, but if they attempt to write to the same bit, only the
processor with the lowest index is allowed to do so.

Then using the equivalence between sequential and parallel growth
models that we showed in the previous section, we have the following:

\begin{proposition}   Given a supply of random bits, a CRCW PRAM with $k$
processors can generate a random internal DLA cluster with $n$
particles in $d$ dimensions in average time $\ord((n/k \,+\, \log k)
\,n^{2/d})$.
\label{practical}
\end{proposition}

\begin{proof}  Using the $k$ processors, we keep $k$ particles active
at any given time, all moving in parallel.  Whenever one or more
reaches an unoccupied site, the particle on the processor with the
lowest index is deposited there, the other particles remain active,
and that processor starts a new particle at the origin.  As we showed
above, this will give us the same probability distribution of clusters
as if we added particles one at a time.

Since each processor adds $n/k$ particles on average, the mean time
for each processor to complete its task is $(n/k)\,\ot$ where $\ot =
\ord(n^{2/d})$.  However, the running time of the algorithm is the
time it takes the last processor to finish, which is at most
$(n/k)\,\ot$ plus the length of the last particle's walk.  Since these
times are distributed with an exponential tail $e^{-t/\ot}$ for large
$t$, and since the average maximum of $k$ things distributed with
probability $P(t) = (1/\ot)\,e^{-t/\ot}$ is
\[ \ot \,\sum_{i=1}^k (1/i) \approx (\gamma + \log k) \,\ot \]
where $\gamma$ is Euler's constant, the last processor finishes in
average time
\[ T = (n/k \,+\, \log k)\,\ot = \ord((n/k \,+\, \log k) \,n^{2/d}) \]
as promised.
\qed \end{proof}

Since the derivative of $n/k \,+\, \log k$ is negative for all $k \le
n$, it pays to add as many processors as we can; for $n$ large enough
that $n \gg k \,\log k$, we get a speedup linear in $k$, which is as
parallelizable as possible.  If we have a massively parallel
computer after all, we can set $k = n$, assign each particle to its
own processor, and get the following corollary:

{\bf Corollary.}
{\em Given a supply of random bits, a CRCW PRAM with $\ord(n)$
processors can generate a random internal DLA cluster of size $n$ in
$d$ dimensions in average time $\ord(n^{2/d} \,\log n)$.}

This corollary gives a middle ground between the algorithm of
Proposition~\ref{jonalg} in the Appendix, which is faster but requires
$\ord(n^{1+2/d})$ processors, and that of Proposition~\ref{practical},
which is slower but requires only a constant number.  This is a nice
example of the tradeoff between computation time and the number of
processors.

\begin{figure}
\centerline{\psfig{file=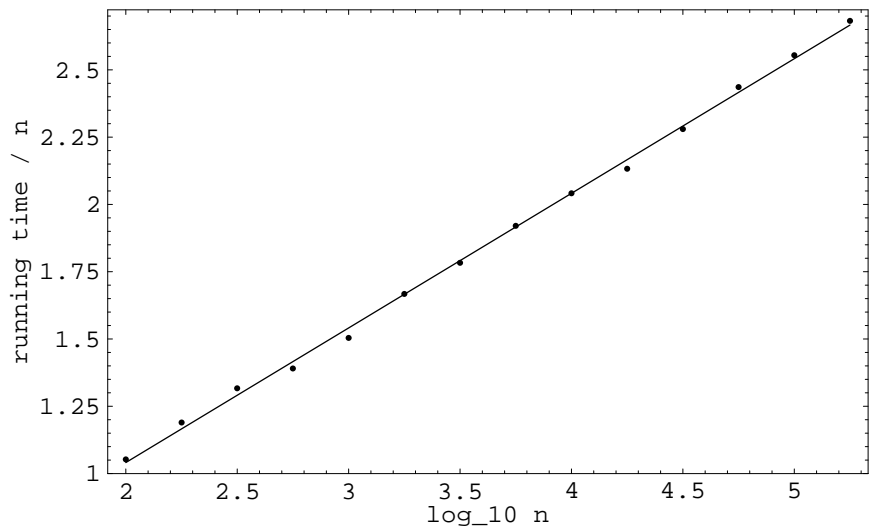,width=4.5in}}
\caption{The running time of the parallel algorithm given by the
Corollary to Proposition~\ref{practical}.  We plot the time divided by
$n$ vs.\ $\log_{10} n$ for $n$ ranging from $10^2$ to $10^{5.25}$ and
averaged over 100 trials each.  Since this is a straight line, the
running time grows as $n \,\log n$.}
\label{running}
\end{figure}

We simulated this algorithm on a serial computer for $d=2$, and found
the same deviations from a circle as in Figure~\ref{dev} within
experimental error.  In Figure~\ref{running} we show the running time
of the algorithm in parallel steps, which is simply the length of the
longest walk.  Taking $100$ trials each for $n$ ranging from $10^2$ to
$10^{5.25}$, we find that the running time does in fact scale as $n
\,\log n$.

\section{An $\NC$ algorithm for internal DLA in $d=1$}

Many problems which are difficult in two or more dimensions are easy
in one.  We will show in this section that a form of {\sc Internal DLA
Prediction} is in $\NC$ for one-dimensional lattices.  This is not
very surprising, since the probability distribution of clusters is
exactly solvable in one dimension \cite{lawler1} but it is good to
have such a result for the record.

\begin{proposition}
{\sc Internal DLA Prediction} on a linear chain with one particle at a
time can be solved by a PRAM with $\ord(n^2)$ processors in $\ord(\log
n)$ time, and so is in $\NC$.
\label{oned}
\end{proposition}

\begin{proof}
In one dimension, the lattice sites are the integers and the cluster
is a line segment $[-L,R]$.  Initially, $L=R=0$.  Each particle's walk
is a mapping that increases either $L$ or $R$ by one, by adding a
particle at the left or right end of the cluster.  Thus, the history
of the cluster can be represented by a directed path starting at the
origin in one quadrant of a two-dimensional square lattice where the
$x$ and $y$ coordinates represent $L$ and $R$ respectively.

For each particle, we have a list of sites it will visit.  The first
step in the algorithm is to convert this list to a mapping on the
square lattice, that is, a table of entries $[L,R] \to [L',R']$ where
$[L',R']$ is either $[L-1,R]$ or $[L,R+1]$, depending on whether $L-1$
or $R+1$ first appears in that particle's list.  Since $L$ and $R$ are
bounded by $n$, this table has length $\ord(n^2)$.  We can do this
conversion in $\ord(\log l)$ parallel time with $\ord(n^2 \,l)$
processors where $l$ is the length of the particle's walk.  If neither
$L$ nor $R$ appears in a particle's list then $[L',R'] = [L,R]$ and
that particle is not incorporated into the cluster.

We then calculate the composition of all these maps by composing the
maps of adjacent pairs of particles, then composing these pairs, and
so on.  This takes $\ord(\log n)$ parallel time and can be done by
$\ord(n^2)$ processors, one for each entry in the map.  The final
state of the cluster is this composed map applied to the initial state
$[0,0]$.  \qed \end{proof}

As a corollary, given a supply of random bits we can generate random
one-dimensional clusters in $\ord(\log n)$ parallel time.  In
addition, the kind of composition process used in the proof can be
carried out by a computer with $\ord(\log n)$ memory, and so is in
$\LOGSPACE \subset \NC^2$ \cite{papa}.

It is interesting that internal DLA on a linear chain can be predicted
in $\NC$ while slight variations of this system are $\CC$-complete.
For example, on a {\em comb} graph, where a linear chain has an
additional site attached to each site on its `backbone,' we can
simulate any comparator circuit as in Proposition~\ref{cccomp} by
using the backbone as our conduit.  (Similarly, by collapsing the
conduit to a single site, we see that internal DLA is also
$\CC$-complete on a {\em star} graph where $n$ sites radiate from a
single central site.)  Nonetheless, from the perspective of
statistical physics, the linear chain and the comb should be in the
same universality class.  For instance, fluctuations in the boundary
should scale as $n^{1/2}$ in both cases.

This situation is familiar from spin glasses, where adding a second
layer to a two-dimensional square lattice changes the problem of
finding the ground state from $\PP$ to $\NP$-complete \cite{barahona},
even though the universality class presumably remains the same.  The
lesson is simply that it is possible to make a problem more difficult
computationally while remaining in the same physical universality
class.

\section{Conclusion}

We have explored the computational complexity of internal
diffusion-limited aggregation.  We have shown that, unlike ordinary
DLA, it cannot make multiple copies of the bits stored on the sites,
and so it is $\CC$-complete rather than $\PP$-complete.  It's pleasing
to find that a ``natural'' problem in physics is complete for a
relatively little-known class of circuits.  We also showed that the
sequential version of the problem is in $\NC$ for a linear chain, even
though it is $\CC$-complete on closely related lattices.

We introduced a dynamic relaxation algorithm in which we guess a
reasonable configuration for the cluster, and then update this witha
non-local annihilation process.  While our numerical measurements are
not definitive, the parallel running time for this algorithm grows
either polylogarithmically in the cluster size $n$ or as a very small
power.  If it is the former, then we have a nice case of a physical
system that can be predicted in $\NC$ on average, even though it is
$\CC$-complete in the worst case.

It is tempting to think that a similar type of algorithm could be of
use in predicting other growth models.  However, since it requires a
number of processors which grows polynomially as a function of system
size, it is unrealistic given the current state of parallel computing.
In the more realistic case where we have a shared-memory computer with
a fixed number $k$ of processors, we used the equivalence between
sequential and parallel growth models and the fact that random
clusters are roughly spherical to show that we can obtain a speedup
which is linear in $k$ for $k \,\log k \ll n$.

{\bf Acknowledgements.}  We are indebted to David Griffeath, Janko
Gravner, Timothy Hely, Erik von Nimwegen, Cosma Shalizi, Aric Hagberg,
Tim Carlson, Frank Gilfeather and Ray Greenlaw for helpful
communications, and to Molly Rose and Spootie the Cat for their
support.  J.M. is upported in part by NSF Grant DMR-9632898.

\appendix
\section{Using $\CC$ to grow random clusters}
\label{app1}

In this section, we show how Mayr and Subramanian's algorithm for
comparator circuits can be used to grow random clusters.  While this
is not the best algorithm, the proof is somewhat instructive.

\begin{proposition}  Given a supply of random bits, for any $\eps > 0$
there is a parallel algorithm that produces a random internal DLA
cluster of size $n$ in $d$ dimensions with probability $1-\eps$ and
runs in $\ord(n \,\log^d (n/\eps) \,\log^2 n)$ parallel time.
\label{ccalg}
\end{proposition}

\begin{proof}  We can convert $td$ random bits into the coordinates of a
$d$-dimensional random walk of $t$ steps in $\ord(\log t)$ parallel
time, since the $j$'th coordinate is the sum of the first $j$ moves.
We then add particles one at a time, by letting our list of moves be a
concatenation of walks, one for each particle.  Note that the
particles will not actually take these walks; they will only take them
as long as they are active, i.e.\ until they reach an unoccupied site.

Since in time $k^{2/d}$ a particle will reach the boundary of a
$d$-dimensional sphere with $k$ sites, the probability of the $k$'th
particle still being active after $t$ steps has an exponential tail of
the form $e^{-t/k^{2/d}}$, and the probability of some particle still
being active at the end of its walk is at most $n$ times this.
Setting this equal to $\eps$ tells us that we can ensure with
probability $1-\eps$ that no particles are left active at time $T$ by
giving the $k$'th particle a walk of length
\[ t = k^{2/d} \,\log (n/\eps) \le n^{2/d} \,\log (n/\eps) \]
Using the construction of Proposition~\ref{cc} gives a comparator
circuit of depth
\[ T < n^{1+2/d} \,\log (n/\eps) \]
and width $n+m$ where $m$ is the total number of sites named in the
walks.  We then use Mayr and Subramanian's simplification algorithm to
evaluate this circuit.

In the worst case where every walk heads away in a different direction
from the origin as fast as it can, $m$ is proportional to $T$, and the
simplification algorithm runs in time $\ord(T \,\log^2 T)$, no better
than explicit simulation.  However, $m$ is almost always significantly
less than $T$, making this circuit narrower than it is deep.  In
particular, since the probability of a particle being at a site a
distance $r$ from the origin after $t$ steps is roughly $t^{-d/2}
\,e^{-r^2/t}$, a crude union bound shows that the probability of any
particle reaching any site $r$ from the origin in $t$ steps is at most
\[ P(r) \,\lesssim\, n \,t^{1-d/2} \,r^{d-1} \,e^{-r^2/t} \]
Setting this equal to $\eps$ tells us that with probability $1-\eps$,
all the particles are confined to a ball of radius
\beq r \,\lesssim\, \sqrt{ t \, \log \frac{n \,t^{1-d/2}}{\eps} }
       \,\lesssim\, n^{1/d} \,\log (n/\eps) \label{bound} \eeq
which is in the crossover regime for multiple random walkers studied
in \cite{larralde}.  The volume of this ball is
\[ m \,\lesssim\, n \,\log^d (n/\eps) \]
and the simplification algorithm works in time
\[ \ord((m+n) \,\log^2 T) \,\lesssim\, n \,\log^d (n/\eps) \,\log^2 n \]
plus smaller corrections.  The two sources of possible error ---
failing to have all the particles' walks terminate, or having some
walker exceed the radius in Equation~\ref{bound} --- both have
probability $\eps$.  By rescaling these to $\eps/2$, we can keep the
total probability of error below $\eps$.
\qed \end{proof}

In fact, this algorithm may run considerably faster, since Mayr and
Subramanian's analysis of their algorithm's running time is based on
the worst-case scenario that each simplification step reduces the
width and depth by only one.  We can expect somewhat better
performance on a random comparator circuit with $N$ gates and width
$W$ whenever $W < N < W^2$.  Since $N=T$, $W \sim m$, and $T \sim
m^{1+2/d}$, this is the case here for $d > 2$.  We leave this
more detailed analysis to the reader.

\section{Shell parallel algorithm}
\label{app2}

In this section, we give a simple parallel algorithm that adds a shell
of constant width to the cluster at each step.  This is equivalent to
sequential or parallel growth by the remarks at the end of
Section~\ref{seccomm}.

\begin{proposition}
Given a supply of random bits, a CRCW PRAM can produce a random
internal DLA cluster of size $n$ in $d$ dimensions in $\ord(n^{1/d}
\,\log n)$ time with $\ord(n^{1+2/d})$ processors.
\label{jonalg}
\end{proposition}

\begin{proof}
First we generate, in advance, the paths of all $n$ walkers; this can
be done in parallel time $\ord(\log n)$ as in the algorithm of
Section~\ref{secrelax}.  We then grow the cluster in a series of
shells.  At each step we take the current cluster $S$ and determine,
in parallel, what site outside $S$ each active particle hits first,
which is where it will stick if no other particle gets there first.
We then look at the set of particles at each sticking point, attach
the one with the lowest index, deactivate it, and keep the other
particles active.  We repeat this with the new cluster, and continue
until there are no active particles left.

In the early stages, the cluster will be diamond-shaped, since almost
every site at its perimeter becomes occupied at each step.  Later on,
it approaches its final shape which is roughly spherical, and every
site on the perimeter has a roughly equal probability of becoming
occupied.  Thus each step adds a shell of constant thickness, and the
algorithm will grow a cluster of size $n$ in $\ord(n^{1/d})$ steps.
Finding the first sticking point of a walk of length $\ord(n^{2/d})$
can be done in $\ord(\log n)$ parallel time with $\ord(n^{2/d})$
processors, so doing this for all $n$ particles takes
$\ord(n^{1+2/d})$ processors.  Finding the particle with the lowest
index at each sticking point can be done in $\ord(\log n)$ time with
just $\ord(n)$ processors.  Therefore, the total running time is
$\ord(n^{1/d} \,\log n)$, and the number of processors we need is
$\ord(n^{1+2/d})$, which is polynomial in $n$.  \qed \end{proof}

This is an adaptation of the parallel algorithm for ordinary DLA given
in Ref.~\cite{moriarty} to internal DLA.  There are two differences
that radically reduce the computation time.  First, in ordinary DLA
particles can block each others' paths, so we have to check for
interactions and throw away all but a non-interacting set.  In
internal DLA, on the other hand, we can treat all the particles in an
almost independent way since the sequential and parallel dynamics are
equivalent, so we can use all the walks at once and none of our
processor time is wasted.  Secondly, the size of an internal DLA
cluster increases linearly with the number of steps since it is
roughly spherical, whereas in ordinary DLA growth is concentrated at
the cluster's protrusions.

\end{document}